\documentclass[twocolumn, pra,showpacs,superscriptaddress]{revtex4}

	\usepackage{amsmath}
	\usepackage{dsfont}
	\usepackage{graphicx}
	\usepackage{epsfig}
	\usepackage[colorlinks,hyperindex]{hyperref}
	\hypersetup
	{
		colorlinks,%
		citecolor=black,%
		linkcolor=black,%
		urlcolor=black,%
	}

\makeindex

\begin{document}

%\title{ Dense Coding with Linear Optics without Bell State Measurements  }
\title{Channel Capacity Gain in Entanglement-Assisted Communication Protocols Based Exclusevly on Linear Optics and Single Photon Inputs }
\author{P. Lougovski } \email{lougovskip@ornl.gov}
\affiliation{Quantum Information Science Group, Oak Ridge National Laboratory, Oak Ridge, TN 37831}
\author{D. B. Uskov} \email{dmitry.uskov@brescia.edu}
\affiliation{Department of Mathematics and Natural Sciences, Brescia University, Owensboro, KY 42301}
\affiliation{Department of Physics and Engineering Physics, Tulane University, New Orleans, LA 70118}

%\date{today}

\begin{abstract}
Entanglement can effectively increase communication channel capacity as evidenced by dense coding that predicts a capacity gain of $1~bit$ when compared to entanglement-free protocols. However, dense coding relies on Bell states and when implemented using photons the capacity gain is bounded by $0.585~bits$ due to one's inability to discriminate between the four optically encoded Bell states. In this paper we study the following question: Are there alternative entanglement-assisted protocols that rely only on linear optics, coincidence photon counting and separable single photon input states and at the same time provide a greater capacity gain than $0.585~bits$. We show that besides the Bell states there is a class of bipartite four-mode two-photon entangled states that facilitate an increase in channel capacity. We also discuss how the proposed scheme can be generalized to the case of two-photon $N$-mode entangled states for $N=6,8$.
\end{abstract}


\maketitle

\section{Introduction}
Dense coding is a quantum communication technique that allows one to send two bits of classical information per single qubit transmitted over a quantum channel. This becomes possible when a sender and a receiver pre-share a maximally entangled two-qubit state (Bell state). Of course, if no entanglement is shared, at most one bit of classical information can be communicated by sending a single qubit. Therefore, using entanglement as a communication resource increases channel capacity by $1~bit$.  The original  protocol~\cite{Bennett} -- closely followed in all optical qubit implementations~\cite{Kwiat1,Weinfurter,DeMartini,Kwiat2} -- was designed for generic qubits and has two main requirements. First, the sender must be able to generate all four Bell states from any given Bell state via a local (single-qubit) operations only. Second, the receiver must be able to perform an unambiguous Bell state discrimination (typically using a two-qubit CNOT gate).  For optical qubits -- most suitable for a long distance communication -- the first requirement can be readily fulfilled by combining spontaneous parametric down conversion entangled photon sources with linear optical devices such as beam splitters, polarization rotators and phase shifters. However, the second requirement of the original protocol is very stringent.  Unfortunately, one  cannot deterministically distinguish all four Bell states only by means of linear optical devices and coincidence measurements~\cite{Luetkenhaus}. Either hyper entanglement~\cite{Weinfurter2,Weinfurter,DeMartini,Kwiat2} or additional entangled ancillae are needed~\cite{Grice,Ewert} making all-optical implementations challenging. As a result, the entanglement-assisted channel capacity gain is bounded by $0.585<1~bits$. But even achieving the reduced bound is fairly difficult due to experimental imperfections. For instance, the actual channel capacity gain reported in~\cite{Kwiat1} was $\approx 0.13~bits$. 

This prompted us to ask the following question: Are there alternative entanglement-assisted protocols (not based on Bell states) that utilize only linear optics, coincidence photon counting and input single photon product states and, at the same time, provide a greater capacity gain than $0.585~bits$? We answer this question by constructing a communication protocol that satisfies the above resource constraints. In particular, we find an equivalence class of four-mode two-photon entangled states that can be prepared using product single photon input states with linear optical elements and transformed into each other by means of linear optical operations on (two) ``local'' modes of one of the parties and, at the same time, can be discriminated by a photon coincidence measurement deterministically. We formalize the problem mathematically in Section~\ref{Sec:Method} and show that the explicit structure of these states as well as the detailed experimental setup that implements our protocol can be obtained by maximizing the mutual information between the sender and the receiver over all physical input states, local/global operations and detection schemes. By solving the optimization task numerically in Section~\ref{Sec:3}, we find that the communication channel capacity of our protocol with ideal detectors is 2 $bits$.
%A distinguishing feature of our protocol is that the sender and the receiver communicate two modes populated with one photon on average and can be %generated using two disentangled single photons and beam splitters only.

To verify that entanglement does indeed provide a gain in the channel capacity we determine the upper bound on the channel capacity with respect to all possible resource-equivalent entanglement-free protocols. We show in Section~\ref{Sec:3} that, under the condition of no vacuum detection, no entanglement-free protocol can achieve the channel capacity greater than 1 $bit$ and, thus, our protocol allows the sender to communicate an extra 1 $bit$ of classical information which is better than the gain of $0.585~bits$ offered by dense coding with linear optics. 

On the other hand, we demonstrate that if vacuum detection is allowed but detectors are imperfect then our protocol still provides a detection-efficiency-dependent channel capacity gain. For example, the gain is $\approx0.27~bits$ for the state-of-the-art superconducting single photon detectors.  Also, our protocol can be extended to the case of two photons shared among $N>4$ modes and in the Section~\ref{Sec:Nmodes} we consider two scenarios with $N=6$ and $N=8$. We show that in both cases solutions exist that provide a capacity gain over corresponding entanglement-free protocols.  

\section{Channel Capacity Formalism for Four-Mode Two-Photon Communication Protocols}\label{Sec:Method}
When implementing an abstract two-qubit system using single photons two so-called ``dual rail'' schemes are prevalent. The first one is the polarization encoding where the logical zero and one states of each qubit are realized as the horizontally $|H\rangle$ and vertically $|V\rangle$ polarized single photon in a given spatial mode and the second is the spatial mode encoding where a single photon placed in either one of two spatial modes  i.e. $|0,1\rangle$ and $|1,0\rangle$ represents the logical zero and one states. The schemes can be mapped onto each other by setting $|H\rangle\equiv|0,1\rangle,|V\rangle\equiv|1,0\rangle$. An arbitrary local (involving modes of one of the parties only) operation can be performed by means of linear optical elements such as a polarization rotator or a beam splitter in the case of dual rail encoding. This class of operations will map the two-qubit computational space $\mathds{C}^{4}=span\{|0,1,0,1\rangle,|0,1,1,0\rangle,|1,0,0,1\rangle,|1,0,1,0\rangle\}$ onto itself. However, when implementing an operation involving spatial modes from both parties with linear optics an input state from $\mathds{C}^{4}$ may end up in a larger Hilbert space $\mathds{C}^{10}$. For example, consider the action of a 50/50 beam splitter between modes 2 and 3 on the state $|\psi\rangle = |0,1\rangle\otimes|1,0\rangle\in\mathds{C}^{4}$. The resulting state $|\tilde{\psi}\rangle = \frac{1}{\sqrt{2}}( |0,0,2,0\rangle-|0,2,0,0\rangle)$ actually lies outside of $\mathds{C}^{4}$ in a larger Hilbert space $\mathds{C}^{10}$. In fact, this becomes an issue when one tries to implement a quantum computer using only linear optical transformations. But for quantum communication problems it may be more advantageous to operate in the full (two photons in four modes) Hilbert space $\mathds{C}^{10}$. 
\begin{figure}[!t]
	\begin{center}
		\includegraphics[scale = 0.45]{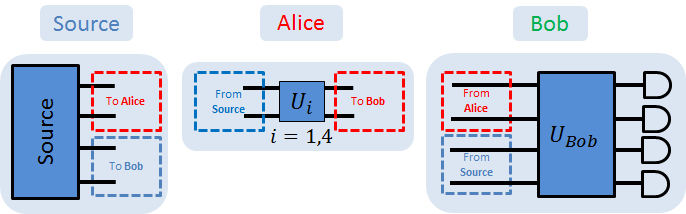}
		\caption{(Color on-line)  Illustration of the proposed entanglement-assisted communication protocol. Alice and Bob share a two-photon four-mode state  $|\Psi_{in}\rangle $ distributed by the source. Alice performs a ``local'' mode transformation $U_{i}$ on her two modes and sends them to Bob who performs a coincidence detection on all four modes and estimates index $i$.}
		\label{Fig:Setup}
	\end{center}
\end{figure}

Indeed, consider the following communication protocol implemented using $N=4$ modes and $n=2$ photons (see Fig.~\ref{Fig:Setup}).  The source generates a special initial two-photon four-mode state $|\Psi_{in}\rangle \in \mathds{C}^{10}$ and sends half of the modes to Alice and the other half to Bob. For the sake of concreteness we assume that Alice gets modes labeled 1 and 2 (1 through $N_{A}$ if $N$ modes are shared). As a result, Alice and Bob share the state  $|\Psi_{in}\rangle$. Upon receiving her part of the state, Alice transforms her modes using one of the four (or potentially more) predetermined {\it two-mode unitary} transformations $U_{i},i=1,\cdots,4$.  She chooses which unitary operation $U_{i}$ to apply  according to a probability distribution $p(U_{i})$ and does not disclose her choice to Bob. Next, Alice sends her part of the state to Bob who now has one of four possible two-photon four-mode states $|{\psi}_{i}\rangle$ with probability $p(|\psi_{i}\rangle)=p(U_{i})$. Bob wants to learn which unitary transformation  $U_{i}$ was performed by Alice i.e. which state $|{\psi}_{i}\rangle$ he has at hand. To do that he sends all four modes through a detection setup that performs a {\it four-mode unitary} transformation $U_{Bob}$ and measures the projection of the output state onto the two-photon four-mode Fock basis states $\{|\phi_{1}\rangle=|2000\rangle,|\phi_{2}\rangle=|1100\rangle,\cdots,|\phi_{9}\rangle=|0011\rangle,|\phi_{10}\rangle=|0002\rangle\}$. Using the outcome of the measurement Bob tries to guess the index $i$ of the state that Alice has sent. 

In qubit-based communication protocols Alice receives a single qubit (i.e. photon) from the source. The photon, depending on the encoding scheme, is distributed over two (polarization or spatial) modes. The local unitary operations $U_{i},i=1,\cdots,4$ that Alice performs on her qubit do not change the total number of photons she communicates to Bob. This provides an implicit constraint on the average number of photons used to communicate a single character i.e. 
\begin{equation}\label{Eq:PhNumConstr}
\sum\limits_{i=1}^{4}p_i\langle \psi_{i}| a_{A1}^{\dagger}a_{A1} + a_{A2}^{\dagger}a_{A2} |{\psi}_{i}\rangle = 1,
\end{equation}   
where $a_{A1}$ and $a_{A2}$ are photon annihilation operators for Alice's mode 1(2) and $p_i$ is the probability that Alice performs an operation $U_{i}$. In our qubit-less protocol the source can potentially provide a state $|\Psi_{in}\rangle$ that violates the constraint in Eq.(\ref{Eq:PhNumConstr}). To make our protocol comparable to dense coding with linear optics  in terms of photonic resources needed to communicate a single character we, therefore, explicitly require that the state provided by the source to Alice and Bob is such that it satisfies the photon number constraint in Eq.(\ref{Eq:PhNumConstr}). 

Naturally, the best case scenario in terms of information transmission(assuming a noiseless quantum channel and perfect detectors) for our protocol is when Alice can prepare four orthogonal states $|{\psi}_{i}\rangle$ subjected to constraints Eq.(\ref{Eq:PhNumConstr}) and Bob can unambiguously detect which state $|{\psi}_{i}\rangle$ Alice has sent him just by means of photon coincidence measurements. In this case, analogously to the original dense coding proposal~\cite{Bennett},  Alice and Bob can communicate two bits of information (provided $p_i = \frac{1}{4} \forall i$) by sending on average one photon in just two modes instead of four. However, whether such states $|{\psi}_{i}\rangle$ (equivalently a state $|\Psi_{in}\rangle$ and two-mode unitaries $U_{i}$) exist is an open question.  Also, to conclude that our protocol is indeed of a dense coding type, we need to determine the largest amount of classical information that Alice and Bob can share by utilizing at most two single photons and two modes under the constraint Eq.(\ref{Eq:PhNumConstr}). Only if the latter is less than $2~bits$ our protocol demonstrates an information gain. To answer these questions we need to formulate the problem in terms of communication channel capacity determination.

From the information-theoretical perspective Alice and Bob form a two-mode two-photon communication system with two auxiliary modes  in which Alice encodes each letter of her message using a two-bit alphabet  $\mathcal{X} = \{|{\psi}_{1}\rangle,\cdots,|{\psi}_{4}\rangle \}$ with probability $p(\psi_{i}) = p(U_{i})$. The receiver, Bob, detects the message sent by Alice as a collection of random signals from the set  $\mathcal{Y} = \{|{\phi}_{1}\rangle,\cdots,|{\phi}_{10}\rangle \}$ with the conditional  probability $p(\mathcal{Y}|\mathcal{X})=p(\psi_{i} | \phi_{j})=|\langle\phi_{i}|\psi_{j}\rangle|^{2}, i=1,\cdots,10;j=1,\cdots,4$ and applies a decoding rule to estimate the original message. Note that $p(\mathcal{Y}|\mathcal{X})$ is a function of the initial state $|\Psi_{in}\rangle$, Alice's unitary operations $U_{i},i=1,\cdots,4$ and Bob's detection setup $U_{Bob}$. In this context, finding the highest rate (in bits) at which information can be sent from Alice to Bob is equivalent to determining information channel capacity of the Alice-Bob system. The information channel capacity $C$ of Alice-Bob channel is defined as~\cite{CoverThomas},
\begin{equation}
\label{Eq:CC}
C = \max I(\psi ; \phi),
\end{equation}  
where the maximization is performed over all possible input distributions $p(\psi_{i})$, states $|\Psi_{in}\rangle$, unitary mode transformations $U_{i}$, $U_{Bob}$ and is subjected to the constraint in Eq.(\ref{Eq:PhNumConstr}); $I$ is the mutual information, 
 \begin{equation}
\label{Eq:MI}
I (\psi ; \phi) = \sum\limits_{j=1}^{4} \sum\limits_{k=1}^{10} p(\psi_{j}, \phi_{k})\log\frac{p(\psi_{j}| \phi_{k})}{p(\psi_{j})}.
\end{equation}
Here $p(\psi_{j}, \phi_{k}) = p(\psi_{j}| \phi_{k})p(\phi_{k})$ denotes the joint probability of Alice preparing the state $|\psi_{j}\rangle$ and Bob detecting the state $|\phi_{k}\rangle$. The marginal probability $p(\phi_{k})$ is defined as $p(\phi_{k}) = \sum_{\psi_{j}} p(\psi_{j}, \phi_{k})$.

By definition, the mutual information $I (\psi ; \phi)$ is a concave function of $p(\psi_{i})$ (3 independent real-valued parameters) and a non-concave function of the  unitary mode transformation matrices $U_{i}\in U(2),i=1,\cdots,4$ (16 real-valued parameters), and $U_{Bob}\in SU(4)$ (15 independent real-valued parameters) and the input state $|\Psi_{in}\rangle$  that we parametrize using ten complex parameters $c_{k}$ (18 independent real-valued parameters~\footnote{only 9 of the coefficients $c_{k}$ are independent because of the normalization constraint $\sum\nolimits_{k=1}^{10}|c_{k}|^2 = 1$ and one of the phases can be set to $0$ }) as $|\Psi_{in}\rangle=\sum\limits_{k=1}^{10}c_{k}|\phi_{k}\rangle$. We can always set one of the matrices $U_{i}$ to be the identity $1\!\!1$ matrix which will leave us with only three independent $2\times2$ unitary matrices. Therefore, the total number of real-valued optimization parameters in Eq.(\ref{Eq:CC}) is 48. 

Since we set Alice's alphabet to only four letters, the global maximum of the channel capacity $C$ over all possible physical setup parameters cannot in principle exceed $\log_{2}(4)=2$ bits. This follows form the definition of the mutual information in Eq.(\ref{Eq:MI}). We observe that for any two random variables $\mathcal{X}$ and $\mathcal{Y}$ : $I(\mathcal{X};\mathcal{Y}) = H(\mathcal{X})- H(\mathcal{X}|\mathcal{Y})$, where $H$ denotes Shannon entropy~\cite{CoverThomas}. Since $H\ge 0$, the maximum of $I(\mathcal{X};\mathcal{Y})$ is achieved when $H(\mathcal{X})$ is maximal ($=\log_{2}(|\mathcal{X}|)$) and $H(\mathcal{X}|\mathcal{Y})$ is minimal (=0) i.e. $\max I(\mathcal{X};\mathcal{Y}) = \log_{2}(|\mathcal{X}|)$. However, it is not clear if this bound is physically attainable. Also, due to the non-concave nature of the optimization objective function many local maximums may exist. Of course, when optimizing $I(\psi ; \phi)$ numerically we are interested in finding a supremum  of all local maximums and hope that it is $2$ bits. Note that because $I$ is concave in parameters $p(\psi_{i})$, if the global ($2$ bits) maximum is attained, using the preceding argument one can immediately show that the only possible values of $p(\psi_{i})=\frac{1}{4}\forall i$. Therefore, we can further reduce the number of real optimization parameters to 45 by setting $p(\psi_{i})=\frac{1}{4}$. 
\begin{table}[!t]
\begin{ruledtabular}
\begin{tabular}{ ccc} 
&Dense&Proposed \\
& Coding & Protocol \\
\hline
Entanglement & Bell States & Multi-mode Entanglement \\ 
\hline
\# of photons  & 1 & 1 \\ 
sent by Alice & & on average\\
\hline
Total \# of photons & 2 & 2\\
\hline
Total \# of modes & 4 & 4\\
\hline
Operations & Linear Optics & Linear Optics \\ 
\hline
Detection & Coincidence & Coincidence
\end{tabular}
\end{ruledtabular}\caption{Resource overview for the proposed entanglement-assisted protocol and conventional dense coding with photonic qubits}\label{T1}
\end{table}
We conclude this section by providing an overview of the resource requirements for our entanglement-assisted communication protocol  as well as for the photon-based dense coding protocol in Table~\ref{T1}. The only difference between the protocols comes from the number of photons that Alice communicates to Bob. In our protocol this number is one on average. Whereas in the dense coding protocol Alice always sends Bob one photon. This requirement also restricts a class of resource-equivalent entanglement-free protocols to those that use one photon on average.

\section{ Four-Mode Two-Photon Protocol Analysis}\label{Sec:3}
\subsection{Optimization Results}
First, to test our approach, we solved the optimization problem in Eq.(\ref{Eq:CC}) using a fixed state $|\Psi_{in}\rangle$ provided by the source. We set $|\Psi_{in}\rangle$ to be equal to one of the Bell states (it does not matter which Bell state is selected, optimization works equally well for all of them) and found by numerical optimization that in this case $C = \log_{2}3$.  Moreover, Alice's mode transformation matrices that correspond to this solution are the same as the ones originally proposed by  Bennett and Wiesner~\cite{Bennett}. It means that by setting the initial state to a Bell state the conventional Bell state-based dense coding protocol~\cite{Bennett,Kwiat1} is recovered.
 
Next, we have discovered, by using gradient-based optimization methods, that the global maximum ($C=2~bits$) is indeed achievable in $\mathds{C}^{10}$. The structure of globally optimal solutions encountered in our numerical search can be parametrized as follows. All globally optimal input states $|\Psi_{in}\rangle$ prepared by the source are, up to a swap of any two modes, equivalent to the state, 
\begin{eqnarray}
\label{Eq:InSt}
|\Psi_{in}\rangle & = & \frac{1}{2}(|1,1,0,0\rangle +|0,1,1,0\rangle +|1,0,0,1\rangle\nonumber \\ 
				   &    & + |0,0,1,1\rangle).
\end{eqnarray}
For example, the following input state
\begin{eqnarray}
|\tilde{\Psi}\rangle & = & \frac{1}{2}(|1,0,1,0\rangle +|0,1,1,0\rangle +|1,0,0,1\rangle\nonumber \\ 
				   &    & + |0,1,0,1\rangle),
\end{eqnarray}
 obtained from $|\Psi_{in}\rangle$ by swapping modes $2$ and $3$ also leads to the globally optimal solution with $C=2~bits$. 
 
Moreover, $|\Psi_{in}\rangle$ in Eq.(\ref{Eq:InSt}) also defines a class of globally optimal input states that are equivalent to $|\Psi_{in}\rangle$ up to a four-mode unitary transformation:
\begin{equation}
\label{Eq:Ut}
 U_{t} = \begin{bmatrix}
U_{A} & 0  \\
0 & U_{B}  \\
\end{bmatrix} ,
\end{equation} 
where $U_{A,B}$ are arbitrary unitary matrices $\in U(2)$, 
\begin{eqnarray}
\label{Eq:UA}
U_{A} & = & \begin{bmatrix}
  e^{i \phi_1} \cos \theta_1 & -e^{i \phi_2} \sin \theta_1 \\
 e^{i \phi_3} \sin\theta_1 & e^{i\left(\phi_2+\phi_3-\phi_1\right)} \cos \theta_1\\
\end{bmatrix}, \\
U_{B} & = & \begin{bmatrix}
  e^{i \phi_4} \cos \theta_2 & -e^{i \phi_5} \sin \theta_2 \\
 e^{i \phi_6} \sin\theta_2 & e^{i\left(\phi_5+\phi_6-\phi_4\right)} \cos \theta_2\\
\end{bmatrix},
\end{eqnarray}
and parameters $\theta_{1,2}$, $\phi_{1,\cdots,6}$ are arbitrary angles $\in[0,2\pi]$. 

Given matrices $U_{A}$ and $U_{B}$, Alice's globally optimal mode transformation matrices (acting on modes 1 and 2) can be decomposed as $ U_{1} = U^{-1}_{A}U_{C}$, 
$ U_{2} = -U^{-1}_{A}\sigma_{z}U_{C}$,  
$ U_{3} = -U_{2}$, $U_{4} = U_{2}\cdot U_{3}$, where $U_{A}$ is defined in Eq.(\ref{Eq:UA}), $U_{C}$ is an arbitrary $2\times 2$ unitary matrix $\in U(2)$ with a similar parametrization and $\sigma_{z}$ denoted Pauli sigma $Z$ matrix.

Lastly, Bob's  four-mode transformation matrix $U_{Bob}$ can be represented as follows,    
\begin{equation}
\label{Eq:UBob}
U_{Bob} = \frac{1}{\sqrt{2}}\begin{bmatrix}
U^{-1}_{C} & 0  \\
0 & U^{-1}_{B}  \\
\end{bmatrix}\begin{bmatrix}
%1 & 0 & 0 & -1 \\
%0 & 1 & -1 & 0\\
%1 & 0 & 0 & 1 \\
%0 & -1 & -1 & 0
1 & 0 & -1 & 0 \\
0 & 1 & 0 & -1\\
1 & 0 & 1 & 0 \\
0 & 1 & 0 & 1
\end{bmatrix}.
\end{equation}
Implementing this mode transformation matrix in an experiment, Bob will detect four distinct coincidence patterns: a coincidence between detectors in modes $1$ and $2$ correspond to Alice's choice $U_{1}$, modes $2$ and $3$ correspond to $U_{2}$, modes $3$ and $4$ correspond to $U_{3}$, modes $1$ and $4$ correspond to $U_{4}$. We remark that this outcome mapping is not unique. Other choices of coincidence assignment are possible and can be realized by additional four-mode unitary rotation on Bob's end i.e. $U_{Bob}\rightarrow U_{Bob}\cdot U_{swap}$. 

\begin{figure}[!t]
	\begin{center}
		\includegraphics[scale = 0.45]{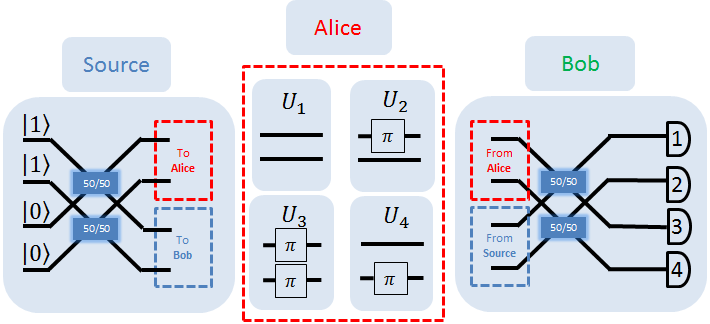}
		\caption{(Color on-line) Linear optical implementation of the proposed entanglement-assisted protocol. Blue rectangles represent $50/50$ beam splitters and transparent squares represent $180^{\circ}$ phase shifters}
		\label{Fig:PhysicalSetup}
	\end{center}
\end{figure}

We emphasize that $U_{t}$, $U_{1},\cdots, U_{4}$ and $U_{Bob}$ define globally optimal unitary {\it mode} transformation. The result of their action on the input state $|\Psi_{in}\rangle$ can be determined in the following fashion. Let us denote input mode photon creation operators  as $a^{\dagger}_{k}, k=1,\cdots,4$, then, 
\begin{equation}
\label{Eq:PolyInState}
|\Psi_{in}\rangle = \frac{1}{2}(a^{\dagger}_{1}a^{\dagger}_{2}+a^{\dagger}_{2}a^{\dagger}_{3}+a^{\dagger}_{1}a^{\dagger}_{4}+a^{\dagger}_{3}a^{\dagger}_{4})|0\rangle,
\end{equation}
where $|0\rangle$ is a four-mode vacuum state. Consider first the unitary mode operation $U_{t}$ defined earlier. It acts on the creation operators $a^{\dagger}_{k}$ as a linear transform i.e. $a^{\dagger}_{k}\rightarrow\sum\limits_{j=1}^{4}(U_{t})_{kj}a^{\dagger}_{j}$, where  $(U_{t})_{kj}$ are the matrix elements of $U_{t}$. As a result, the input state $|\Psi_{in}\rangle$ in Eq.(\ref{Eq:PolyInState}) is transformed to
\begin{eqnarray}
\label{Eq:PolyInStateTranformed}
|\Psi_{t}\rangle & = &\frac{1}{2}(\sum\limits_{i,j=1}^{4}[(U_{t})_{1i}(U_{t})_{2j}+(U_{t})_{2i}(U_{t})_{3j}+(U_{t})_{1i}(U_{t})_{4j} \nonumber\\
& & + (U_{t})_{3i}(U_{t})_{4j}]a^{\dagger}_{i}a^{\dagger}_{j})|0\rangle.
\end{eqnarray}
Similarly, unitary mode transformations  $U_{1},\cdots, U_{4}$ and $U_{Bob}$ can now be applied to the state $|\Psi_{t}\rangle$ in Eq.(\ref{Eq:PolyInStateTranformed}) in sequence, resulting in the desired output state at Bob's detectors.

The structure of the globally optimal solution is most transparent when $U_{A}=U_{B}=U_{C}=1\!\!1$. In this case the physical setup that implements our communication protocol is illustrated in Fig.(\ref{Fig:PhysicalSetup}). Surprisingly, its structure is equivalent to a double Mach-Zehnder interferometer. The source prepares the state $|\Psi_{in}\rangle$ in Eq.(\ref{Eq:InSt}) by placing a separable state containing two single photons in modes $1$ and $2$ i.e. $|1,1,0,0\rangle$ onto two $50/50$ beam splitters coupling modes ($1$,$3$) and ($2$,$4$).  To transform her two modes, Alice needs only $180^{\circ}$ phase shifters. Note that the transformation $U_{3}$ applies a phase shift to both Alice's modes with respect to Bob's modes. Bob recombines modes ($1$,$3$) and ($2$,$4$) on two $50/50$ beam splitters and sends them for the coincidence detection. In the case of noiseless channel and perfect detectors this converts the states $|\psi_{1}\rangle\rightarrow|1,1,0,0\rangle$, $|\psi_{2}\rangle\rightarrow|0,1,1,0\rangle$, $|\psi_{3}\rangle\rightarrow|0,0,1,1\rangle$, $|\psi_{4}\rangle\rightarrow|1,0,0,1\rangle$ which results in the channel capacity $C = 2~bits$. Also note that photon number resolving detectors are not required by Bob since he relies on a coincidence detection.

\subsection{Channel Capacity of Resource-Equivalent Entanglement-Free Protocols}
Let us now calculate the channel capacity for the case when Alice and Bob do not share any entanglement in order to determine whether our entanglement-assisted protocol provides any advantage for information transmission purposes. For that we consider the following resource-equivalent scenario: Alice prepares a {\it two-mode} state $|\tilde{\psi}_{i}\rangle, i = 1,\cdots,M$, corresponding to a character from a $M$-letter alphabet $\mathcal{X}$ ($M$ is to be determined later), with probability $p(\tilde{\psi}_{i})$ and sends it to Bob who can perform an arbitrary {\it two-mode} unitary transformation $U_{Bob2}$ before he detects the received state using two photo detectors that are {\it not} number resolving (because the correspondent entanglement-based protocol does not require photon-number-resolving detectors).

To facilitate a fair comparison between the entanglement-free and entanglement-assisted protocols, we assume that Alice's resources per character are limited to at most {\it two single photons} and she may use only linear optical elements(i.e. only two-mode unitary operations) to prepare states $|\tilde{\psi}_{i}\rangle$. In this case, inputs to Alice's state preparation setup are limited to $|0,0\rangle, |0,1\rangle,|1,0\rangle,|1,1\rangle$. Linear optical operations that Alice may perform to prepare states $|\tilde{\psi}_{i}\rangle$ conserve the total number of photons and, thus, $|\tilde{\psi}_{i}\rangle$ cannot be a superposition of states with different total photon numbers such as, for instance,  $|0,0\rangle+|1,1\rangle$ or  $|0,0\rangle+|0,1\rangle$, etc. Also, the states $|\tilde{\psi}_{i}\rangle$ must be mutually orthogonal. If they are not, Bob will be forced to discriminate between non-orthogonal states since he can only apply a two-mode unitary transformation (unitary transformations do not change the overlap between the states $|\tilde{\psi}_{i}\rangle$) to all states he receives. This will effectively reduce the amount  of information that Bob and Alice can exchange and make such a communication protocol suboptimal. 

The subspaces that contain states with different total number of photons are orthogonal and we only need to determine how many orthogonal states can be created by Alice in each subspace by means of a two-mode unitary transform. The subspace spanned by vacuum $|0,0\rangle$ is invariant under two-mode unitary transforms and Alice can use  $|0,0\rangle$ as $|\tilde{\psi}_{1}\rangle$. The one-photon subspace spanned by $|0,1\rangle$ and $|1,0\rangle$ trivially contains two orthogonal states e.g. $|\tilde{\psi}_{2,3}\rangle = \frac{1}{\sqrt{2}}(|0,1\rangle \pm |1,0\rangle)$ that can be generated using a 50/50 beamsplitter. Finally, it is straightforward to show that only one state orthogonal to $|1,1\rangle$ can be generated by two-mode unitary operations in the two-photon subspace. Hence, Alice can choose $|\tilde{\psi}_{4}\rangle = \frac{1}{\sqrt{2}}(|2,0\rangle-|0,2\rangle)$ and $|\tilde{\psi}_{5}\rangle=|1,1\rangle$. This implies that Alice may prepare up to $M=5$ mutually orthogonal states by using linear optical elements. 

We now can calculate the channel capacity for the above choice of the states $|\tilde{\psi}_{i}\rangle, i=1,5$, maximizing the mutual information function over Bob's unitary transformation $U_{Bob2}$ and probabilities $p(|\tilde{\psi}_{i}\rangle)$, subjected to the average photon number constraint in Eq.(\ref{Eq:PhNumConstr}) which for this case reads, 
\begin{equation}\label{Eq:MeanPhConstr}
p_{2}+p_{3}+2p_{4}+2p_{5}=1.
\end{equation} 
We found, by solving the optimization task numerically, that Bob's optimal transformation is a 50/50 beam splitter and the channel capacity in this case amounts to 2 $bits$. The optimal values for $p_{i}$'s are $p_1=p_2=p_3=p_5=0.25;~p_4=0$. Therefore, Alice can equivalently (up to a two-mode unitary transformation) use the following set of states to encode her message: 
\begin{equation}\label{Eq:Aset}
 \{|\tilde{\psi}_{i}\rangle\}=\{|0,0\rangle,|0,1\rangle,|1,0\rangle,|1,1\rangle\}.
 \end{equation} 
This is because Bob's detectors are not number resolving, and he cannot discriminate a single photon event from a two-photon event. Therefore, even if Alice can prepare five mutually orthogonal states the amount of information that Bob can extract from them is the same as if Alice used the four state alphabet in Eq.(\ref{Eq:Aset}).   

At first sight one may conclude that if Bob has ideal(100$\%$ efficient) detectors and additional synchronization information to distinguish the vacuum signal $|0,0\rangle$ from no signal events, he could discriminate the states $|\tilde{\psi}_{i}\rangle$ in Eq.(\ref{Eq:Aset}) perfectly which implies that the channel capacity of the entanglement-free protocol is also 2 $bits$. However, if the vacuum detection is ruled out, it would further reduce the set $\{|\tilde{\psi}_{i}\rangle\}$ to only three states i.e.  $\{|\tilde{\psi}_{i}\rangle\}=\{|0,1\rangle,|1,0\rangle,|1,1\rangle\}$. Moreover, by combining Eq.(\ref{Eq:MeanPhConstr}) with the normalization condition $\sum\limits_{i=1}^{5}p_{i}=1$ and setting $p_1=p_4=0$ we immediately derive that $p_5=0$. So effectively this becomes a two-mode two-state($\{|\tilde{\psi}_{i}\rangle\}=\{|0,1\rangle,|1,0\rangle\}$) protocol with the channel capacity of $1~bit$. Therefore, under the condition of no vacuum detection, our entanglement-based protocol allows one to communicate one extra bit of information when compared to the best possible entanglement-free two-mode communication scheme.

Lastly, let us consider what happens when the vacuum detection is allowed but Bob's detectors are non-ideal. We show in Appendix~\ref{Appendix} that in this case the difference in the channel capacity of the proposed entanglement-based protocol and its resource-equivalent entanglement-free analogue discussed earlier in this subsection is a positive function of the detection efficiency (see Fig.(\ref{Fig:ChannelCapacityDifference}) for details). Thus, we show that for realistic detectors our entanglement-based protocol always allows transmission of more information then its entanglement-free version with the same amount of resources used.

\subsection{Channel Capacity as a Function of Average Photon Number}
Next, let us study how much information can be communicated in our entanglement-assisted protocol by sending less than one photon on average. This can be readily done by modifying the optimization constraint in Eq.(\ref{Eq:PhNumConstr}). Note that, since Alice only uses passive optical elements, the average number of photons she sends to Bob is actually controlled by the source i.e. 
\begin{equation}
\langle N_{Alice}\rangle = \langle\Psi_{in}|a^{\dagger}_{A1}a_{A1}+a^{\dagger}_{A2}a_{A2}|\Psi_{in}\rangle,
\end{equation} 
where $a_{A1,A2}$ denote photon annihilation operators for Alice's modes $1,2$. Thus, the new optimization problem at hand is, 
\begin{eqnarray}
\label{Eq:CCplusConstraint}
 maximize~I(\psi ; \phi) \nonumber \\
 s.t. ~\langle N_{Alice}\rangle = n,
\end{eqnarray}
\begin{figure}[!t]
	\begin{center}
		\includegraphics[width=0.5\textwidth]{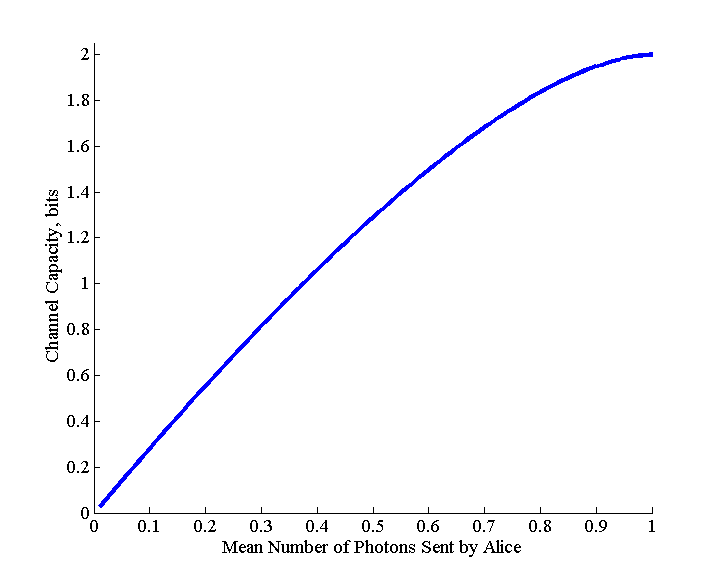}
		\caption{(Color on-line) Channel capacity as a function of the mean number of photons communicated by Alice. }
		\label{Fig:lCapacityVsPhotons}
	\end{center}
\end{figure}   
where $n$ is a constant $\in [0,1]$. We used a gradient-based solver to optimize Eq.(\ref{Eq:CCplusConstraint}) numerically for various values of the average photon number $n$. For a fixed value of $n=0.05$ we ran 500 independent optimizations using random starting points. We then selected the largest value of $\max I(\psi ; \phi)$ over these runs and used it as a starting point for optimizing $I(\psi ; \phi)$ for the next value of $n=0.1$. By gradually varying the constraint value we  calculated $\max I(\psi ; \phi)$ as a function of $n$ depicted in Fig.(\ref{Fig:lCapacityVsPhotons}). We observe that if Alice sends Bob less then one photon on average then their channel capacity falls below two bits. However, they can achieve channel capacity $C \approx 1.63~bits$ (the best value demonstrated to date~\cite{Kwiat2}) by communicating $\approx 0.68 < 1$ photons on average.  

\subsection{Communicating Larger Alphabets}
\begin{figure}[!t]
	\begin{center}
		\includegraphics[width=0.5\textwidth]{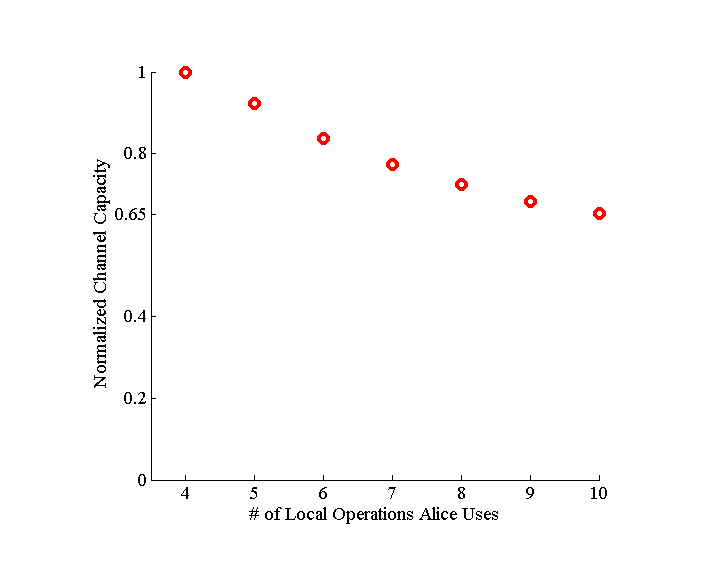} 
		\caption{(Color on-line) Normalized channel capacity. Red circles denote the channel capacity of our entanglement-assisted protocol obtained by solving Eq.(\ref{Eq:CC}) as a function of the number of Alice's local operations $M$ normalized to the theoretical maximum of $\log_{2}M$.}
		\label{Fig:ChannelCapacity}
	\end{center}
\end{figure}

In principle, Alice may want to use larger alphabets than just the four-symbol one. After all, we are operating with the states in $\mathds{C}^{10}$ and naturally the question arises whether a physical setup exists that attains the channel capacity $C=\log_{2}M$ for some integer $M\in[5,10]$. To answer this question we modified our protocol by allowing Alice to perform $M>4$ unitary transformations on her two modes. At the same time we still require Bob to measure in the Fock basis $\{|\phi_{j}\rangle\}, j=1,10$. We numerically optimized the channel capacity in Eq.(\ref{Eq:CC}) for the cases of $M=5,\cdots,10$ and normalized the respective values to the maximal theoretically attainable channel capacity ($C=\log_{2}M$). The results are plotted in Fig.(\ref{Fig:ChannelCapacity}). We notice that the maximal theoretical channel capacity is only achievable for the case of the two-bit alphabet (M=4). When Alice is trying to use $M>4$ symbols in her alphabet the normalized channel capacity decreases. This is because Bob is unable to deterministically discriminate between the states $|\psi_{i}\rangle, i=1,\cdots,M; M>4$ by using projective measurements in Fock basis.      

\section{Optimization Results For $N=6,8$ Modes $n=2$ Photons States}\label{Sec:Nmodes}
 In principle, the protocol described in Sec.~\ref{Sec:Method} can be used for linear optical circuits with arbitrary number of modes $N$ and photons $n$. Here, we study two simple extensions with two photons ($n=2$) distributed over $N=6$ and $N=8$ modes. We assume that in both cases Alice and Bob receive $N/2$ modes from the source. Our goal is again to solve numerically the channel capacity problem posed in Eq.(\ref{Eq:CC}). 

For the $n=2$, $N=6$ case the dimensionality of the correspondent Hilbert space is $\dim\mathds{C} = \frac{(n+N-1)!}{n!(N-1)!}=21$ which naturally leads to the question: is it possible to design an entanglement-assisted communication scheme that provides channel capacity of $\log_{2}21\approx 4.39~bits$ by sending just three modes from Alice to Bob? The necessary condition for that is Alice must be able to prepare $21$ orthogonal states from an input state $|\Psi_{in}\rangle$ by means of ``local'' three-mode unitary transformations. However, we discovered numerically that Alice can at best prepare $12$ orthogonal states using  three-mode unitary transformations. This means that the channel capacity cannot  possibly exceed $\log_{2}12\approx 3.58~bits$. Next, we numerically solved the optimization task in a modified version of Eq.(\ref{Eq:CC}) for the case of $M=12$ local $U(3)$ operations performed by Alice and one global $U(6)$ mode transformation performed by Bob. Obtained solutions suggest $C = 3.0~bits$ which implies that even if Alice can locally prepare 12 orthogonal states, Bob cannot discriminate then deterministically by means of linear optics and coincidence detection. Indeed, we discovered that Bob can only perform a non-ambiguous detection of 8 orthogonal states encoded by Alice. Therefore, the practical(achievable) channel capacity in the case of $n=2$ photons in $N=6$ modes is limited to $3~bits$. This result was obtained using an unconstrained optimization of the mutual information function. The actual mean number of photons per character in this scheme is $1.5$ photons. Correspondent entanglement-free three-mode communication schemes with the same photon-per-character cost are equivalent to a six state protocol where Alice sends states $\{|1,0,0\rangle,|0,1,0\rangle,|0,0,1\rangle,|1,1,0\rangle|1,0,1\rangle,|0,1,1\rangle\}$ which results in the channel capacity of $\log_{2}6\approx 2.585~bits$(assuming perfect detectors and no vacuum detection constraint). Therefore, the information gain in the  $n=2$ photon, $N=6$ modes dense coding scheme is  $\approx 0.415<1~bits$.

Similar analysis for the case of $n=2$ photons in $N=8$ modes revealed that the practical channel capacity of the eight-mode communication is limited to $\log_{2}12\approx 3.58~bits$ (compare to the four-mode entanglement-free channel capacity of $\log_{2}10\approx 2.3~bits$).
\section{Summary}
We discussed the problem of entanglement-assisted communication channel capacity gain for a linear optical circuit with $N$ modes populated by $n$ photons. We discovered in the case of $N=4$, $n=2$ there is a class of mode-entangled states that supports protocols with a range of channel capacity gain over corresponding resource-equivalent entanglement-free protocols.  We studied numerically 6 and 8 mode extensions of the protocol and provided estimates for the channel capacity in those cases. 
\begin{acknowledgements}
This work was performed at Oak Ridge National Laboratory, operated by UT-Battelle for the U.S. Department of energy under contract no. DE-AC05-00OR22725. PL would like to thank Warren P. Grice for useful discussion. DBU acknowledges support from the NSF under grant PHY-1005709. 
\end{acknowledgements}
\appendix 
\section{Channel Capacity with Imperfect Detectors}\label{Appendix}
First, let us discuss why the detection of vacuum with a photon-counting detector is less efficient than the detection of a photon with the same detector. We assume that for any input state of light the detector produces either a ``click'' outcome (denoted as $+$) or a ``no click''  outcome (denoted as $-$). Next, we denote $p(-|0) = v$ the conditional  probability of  the detector producing a ``no click'' outcome, given the input state on the detector was vacuum $|0\rangle$. We also introduce the conditional probability $p(+|1) = s$ of the detector clicking provided the input state was a single photon $|1\rangle$.  In a similar fashion we define the conditional probability $p(+|0) = 1-v$ for the detector to click erroneously when the input state was  $|0\rangle$ and the probability $p(-|1) = 1 - s$ for the ``no click'' event when the input contained a photon. Note that for imperfect detectors ``no click''(``click'') does not necessary imply that the input state was zero(one) photon.

Assuming that states $|0\rangle$ and $|1\rangle$ have equal probabilities of arriving at the detector i.e. $p(0)=p(1)=\frac{1}{2}$ let us calculate the probability $p(0|-)$ of the vacuum state arriving at the detector provided a ''no click'' event was recorded. Using Bayes' rule we obtain,
\begin{equation}\label{Eq:VacEff}
p(0|-) =  \frac{p(-|0)\cdot p(0)}{p_{NC}} =  \frac{v}{1-s+v},
\end{equation}
where the total probability of a``no click'' event $p_{NC}$ is given by
\begin{equation}
p_{NC} = p(-|0)\cdot p(0) + p(-|1)\cdot p(1) = \frac{1-s+v}{2}.
\end{equation}
Similarly, we can compute the probability $p(1|+)$ of a single photon arriving at the detector provided a ''click'' event was recorded as
\begin{equation}\label{Eq:SPEff}
p(1|+) = \frac{p(C||0\rangle)\cdot p(|0\rangle)}{p_C} = \frac{s}{1+s-v},
\end{equation}
where the total probability of a ''click'' event $p_C$ is given by
\begin{equation}
p_C = p(+|0)\cdot p(0) + p(+|1)\cdot p(1) = \frac{1+s-v}{2}.
\end{equation}
Comparing Eq.(\ref{Eq:VacEff}) and Eq.(\ref{Eq:SPEff}) for a fixed value of $v$ ($v\approx1$) we observe that $p(1|+)>p(0|-)$ for any finite efficiency detector ($s<1$).  For a typical state-of-the-art superconducting single photon detector~\cite{Marsili} $s\le 0.9$ and $1-v\approx 10^{-4}$ which translates into the vacuum detection efficiency of $p(0|-)\approx 0.91$. On the other, hand the efficiency of a single photon detection is $p(1|+) = 0.9999$. Therefore, detecting a single photon in a single mode is almost $10\%$ more efficient than detecting vacuum. Furthermore, for the two-mode vacuum state $|0,0\rangle$ the detection efficiency is $\approx 0.82$ which is significantly less than the detection efficiency for states $|1,1\rangle$ ($\approx 0.9998$) and $|0,1\rangle$ ($\approx 0.91$). Such a disparity in detection efficiency suggests that linear optical schemes that rely on double vacuum detection will experience a more severe channel capacity reduction than coincidence-based schemes.

With the previous discussion in mind, let us now quantify the effect of imperfect detectors onto the achievable channel capacity for both entanglement-free and entanglement-assisted communication protocols proposed in Section~\ref{Sec:3}. In the entanglement-free case, Bob uses two imperfect photon detectors to resolve between four possible states in Alice's alphabet $\mathcal{X}=\{|00\rangle,|01\rangle,|10\rangle,|11\rangle\}$.  Therefore, Bob's alphabet contains four possible detection outcomes $\mathcal{Y}=\{--,-+,+-,++\}$. Assuming that Bob's detectors have the same efficiency $s<1, v\approx1$ we can calculate the mutual information function $I(\mathcal{X};\mathcal{Y})$  between Alice and Bob as,
\begin{equation}\label{Eq:MI2modes}
I (\mathcal{X};\mathcal{Y}) = \sum\limits_{j=1}^{4} \sum\limits_{k=1}^{4} p(x_{j}, y_{k})\log\frac{p(x_{j}, y_{k})}{p(x_{j})p(y_{k})},
\end{equation}    
where the probability $p(x_{j}, y_{k})$ can be expressed in the matrix form,
\begin{equation}\label{Eq:JP1}
%p(\mathcal{X};\mathcal{Y}) = 
\begin{bmatrix}
v^2p_1 & v(1-v)p_1 & v(1-v)p_1 & (1-v)^2p_1 \\
v(1-s)p_2 & vsp_2 & (1-s)(1-v)p_2 & s(1-v)p_2\\
v(1-s)p_3 & (1-v)(1-s)p_3 & svp_3 & s(1-v)p_3 \\
(1-s)^2p_4 & s(1-s)p_4 & s(1-s)p_4 & s^2p_4\\
\end{bmatrix}
\end{equation}
\begin{figure}[!t]
	\begin{center}
		\includegraphics[width=0.5\textwidth]{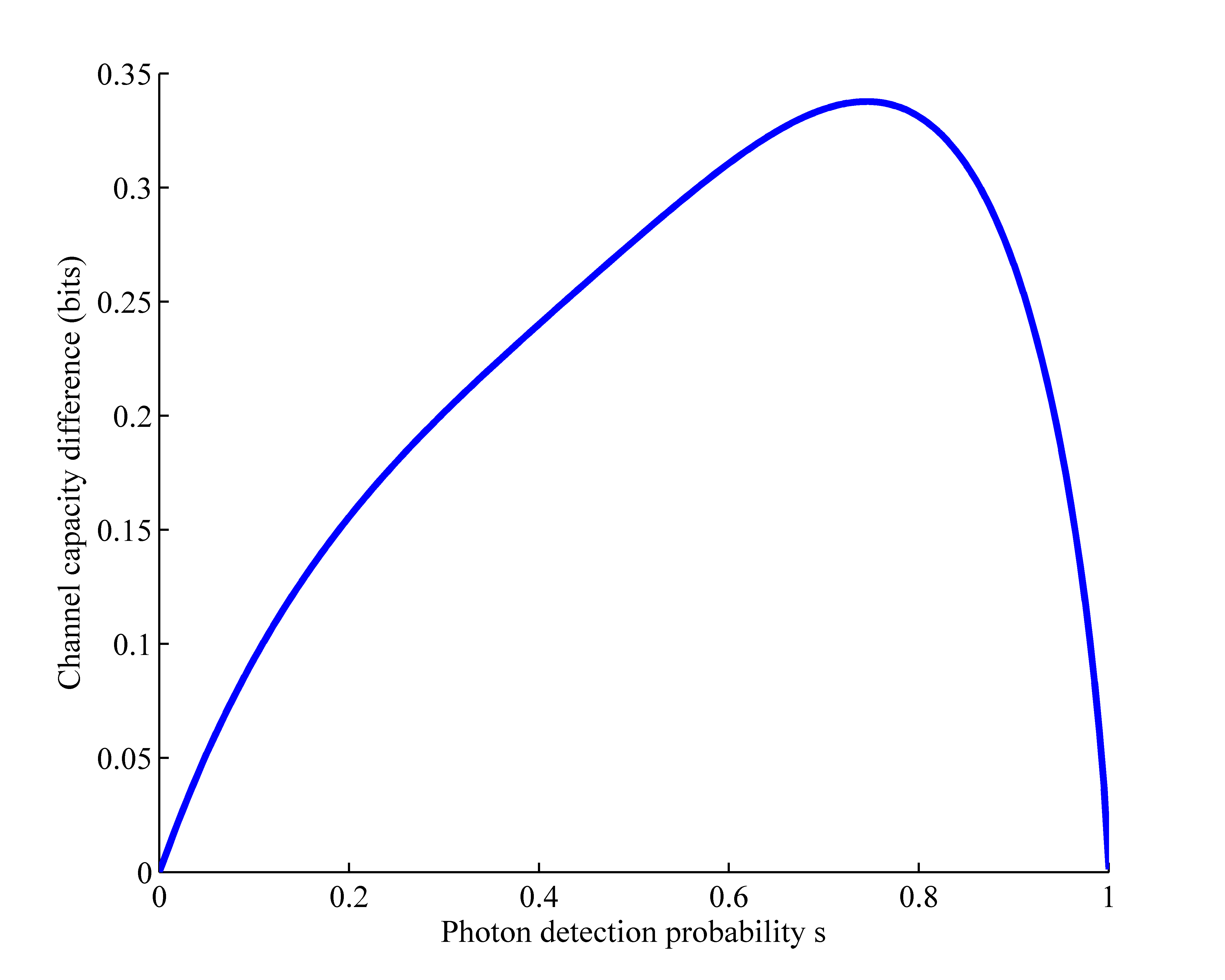} 
		\caption{(Color on-line) Channel capacity difference between entanglement-assisted and entanglement-free communication protocols as a function of the photon detection probability s. }
		\label{Fig:ChannelCapacityDifference}
	\end{center}
\end{figure}
with the entry in the $j$-th row and $k$-th column representing a joint probability of Alice sending a state $x_j$ from $\mathcal{X}$ and Bob detecting an outcome $y_k$ from $\mathcal{Y}$. Here $p_i, i=1,4$ is the probability of Alice sending the $i$-th state from her alphabet i.e. $p_i=p(x_{i})\forall i$. An explicit expression for the mutual information can be  readily obtained by combining Eqs.(\ref{Eq:MI2modes}-\ref{Eq:JP1}). For fixed values of $s$ and $v$, $I (\mathcal{X};\mathcal{Y})$ is a function of only two independent parameters e.g. $p_1$ and $p_2$ ($p_3$ and $p_4$ can be expressed in terms of  $p_1$ and $p_2$ using Eq.(\ref{Eq:MeanPhConstr}) and probability normalization) and maximizing it numerically gives the channel capacity $C_{noent}$ for the entanglement-free communication scheme.  In a similar fashion we can compute the channel capacity for the entanglement-assisted protocol. In this case Alice prepares one of the four states from the set $\mathcal{X}=\{|\psi_1\rangle,\cdots,|\psi_4\rangle\}$ defined in Section~\ref{Sec:Method} and Bob registers one of 16 possible four-mode click patterns $\mathcal{Y}=\{----,+---,\cdots,++++\}$. The mutual information $I(\mathcal{X};\mathcal{Y})$  can now be calculated by extending the summation range over $k$ from 4 to 16 in Eq.(\ref{Eq:MI2modes})  and introducing a correspondent $4\times 16$ join probability matrix $p(x_{j}, y_{k})$. Note that unlike in the entanglement-free protocol now $I(\mathcal{X};\mathcal{Y})$ depends on three parameters $p_1,\cdots,p_3$  for a fixed value of $s$ and $v$. The final algebraic expression can be optimized numerically to find the channel capacity of the entanglement-assisted protocol $C_{ent}$. 

We have obtained values of $\Delta C = C_{ent} - C_{noent}$ numerically by using a random search method on a set of $10^7$ randomly generated probability distributions $\{p_i\}$ for various values of the single photon detection probability $s$ and $v=0.9999$. The results are displayed on Fig.(\ref{Fig:ChannelCapacityDifference}). We noticed that in the limit of perfect detectors both protocols reach the same channel capacity ($2~bits$). However, with realistic detectors ($s<1$) the entanglement-assisted protocol exhibits an information gain over its entanglement-free counterpart.  In particular, for the best superconducting detectors ($s\approx0.9$) the information gain is $\approx 0.27~bits$. Although, the gain is $<1~bit$, it is greater than the actual experimental gain of $0.13~bits$ observed in~\cite{Kwiat1}.

\end{document}